\documentclass[a4paper,11pt]{article}
\pdfoutput=1 

\usepackage{graphicx}
\usepackage{color}
\usepackage{amsfonts,amssymb,latexsym,amsmath}
\usepackage{a4wide}
\usepackage{bm}
\usepackage{array} 
\usepackage{hyperref}
\usepackage{epstopdf}
\usepackage{cite}

\newcommand{\be}{\begin{equation}}
\newcommand{\ee}{\end{equation}}
\newcommand{\ba}{\begin{eqnarray}}
\newcommand{\ea}{\end{eqnarray}}

\newcommand{\al}{&\!\!\!}

\newcommand{\sci}{\mathcal{I}}

\newcommand{\deltaL}[1]{\delta_L\!\left[#1\right]}

\begin{document}

\title{\bf Finite-volume corrections to the CP-odd nucleon matrix
elements of the electromagnetic current from the QCD vacuum angle}

\author{
Tarik Akan$^{a,b}$\thanks{tarik.akan@bozok.edu.tr},~~
Feng-Kun Guo$^{b}$\thanks{fkguo@hiskp.uni-bonn.de}~~
and Ulf-G. Mei{\ss}ner$^{b,c}$\thanks{meissner@hiskp.uni-bonn.de}\\
{\small \it $^a$Physics Department, Bozok University,  66200 Yozgat, Turkey }\\
{\small \it $^b$Helmholtz-Institut f\"ur Strahlen- und Kernphysik and Bethe
   Center for Theoretical Physics,}\\
{\small \it Universit\"at Bonn,  D-53115 Bonn, Germany}\\
{\small\it $^c$Institute for Advanced Simulation, Institut f\"{u}r Kernphysik
   and J\"ulich Center for Hadron Physics,}\\
{\small\it JARA-FAME and JARA-HPC, Forschungszentrum J\"{u}lich,
   D-52425 J\"{u}lich, Germany}
}

\maketitle

\begin{abstract}
Nucleon electric dipole moments originating from
strong CP-violation are being  calculated by several groups using
lattice QCD. We revisit the finite volume corrections to the CP-odd nucleon
matrix elements of the electromagnetic current, which can be related to the
electric dipole moments in the continuum, in the framework of
chiral perturbation theory up to next-to-leading order taking into account the
breaking of Lorentz symmetry. A chiral extrapolation of the recent lattice
results of both the neutron and proton electric dipole moments is performed,
which results in $d_n=(-2.7\pm1.2)\times10^{-16}e\,\theta_0\,$cm and
$d_p=(2.1\pm1.2)\times10^{-16}e\,\theta_0\,$cm.
\end{abstract}

\newpage

\section{Introduction}

The electric dipole moment (EDM) measures the polarity of a system of charged
particles. For a hadron or any other elementary particle at rest, it can be
expressed as $\vec{d} = d\, \hat{{S}}$, where $\hat{{S}}$ is the
direction of the spin of the particle, and the coefficient $d$ refers to the
EDM that will be discussed in this paper. The interaction of a dipole with an
electric field $\vec{E}$ is described by the Hamiltonian
$\mathcal{H}_\text{edm} = - \vec{d}\cdot\vec{E} $. Such an interaction changes
its sign under both parity transformation (P) or a combined transformation of
parity and charge conjugation (C). Thus, the nucleon EDM is a CP-odd quantity.
In the Standard Model (SM), the Cabibbo--Kobayashi--Maskawa (CKM) contribution
to the neutron EDM is tiny, which was estimated to be $1.4\times10^{-33}\leq
|d_n|\leq 1.6\times10^{-31}~e\,$cm~\cite{He:1989xj}. Thus, the nucleon EDM
serves as a sensitive
probe of  physics beyond the SM. So far, there has not been
an experimental evidence for a non-zero nucleon EDM, and the current experimental
upper bound for the neutron EDM is $|d_n| \leq 2.9 \times 10^{-26}\, e
\,$cm~\cite{Baker:2006ts}, which is several orders of magnitude larger than
the value from the CKM mechanism. For the proton EDM, the experimental upper
bound as derived from the EDM of the $^{199}$Hg atom is
$|d_p|<7.9\times10^{-25}e\,$cm~\cite{Griffith:2009zz}.

Within the SM, in  addition to the CKM mechanism, there is another
source of CP violation, which is from the $\theta$-term of quantum
chromodynamics (QCD). Therefore, the experimental information on the nucleon
EDMs allows us to constrain both physics beyond the SM (BSM) models and the
value of the QCD vacuum angle $\theta_0$. For a recent review on the EDMs of nucleons,
nuclei and atoms both within and beyond the SM, we refer to
Ref.~\cite{Engel:2013lsa}.

Besides the active and planned experimental activities~(for a brief review,
see Chapter~7.2 of Ref.~\cite{Hewett:2012ns}, and a collection of various
experiments  can be found on the webpage~\cite{EDMexps}), the contribution of
the $\theta$-term to the nucleon EDMs is being calculated using lattice
QCD~~\cite{Aoki:1989rx,Aoki:1990ix,Shintani:2006xr,Shintani:2008nt,
Shintani:2005xg,Berruto:2005hg,Izubuchi:2008mu,Aoki:2008gv,Schierholz:2013talk,
Shintani:2014talk }.
All these lattice calculations were performed with  up and down quark masses
larger than their physical values, or equivalently with a pion mass
larger than its physical mass. One of the methods used in the
lattice calculation is to
calculate the CP-odd electric dipole form factor (EDFF) of the nucleon
$F_{3,N}(q^2)$ in the space-like region with finite $-q^2$. The definition of
the nucleon EDFF in the infinite volume is given by
\begin{eqnarray}
  \left\langle N(p',s') | J^\nu | N(p,s) \right\rangle \al=\al
   \frac{F_{3,N}\left(q^2\right)}{2m_N}
   \bar u(p',s') \sigma^{\mu\nu}q_\mu\gamma_5  u(p,s)
   + \ldots \nonumber\\
   \al=\al i \frac{F_{3,N}\left(q^2\right)}{2m_N}  \left(p+p'\right)^\nu
   \bar u(p',s') \gamma_5 u(p,s) + \dots~,
   \label{eq:formfactor}
\end{eqnarray}
with $J^\nu$ being the electromagnetic current, $p$ and $p'=p+q$ the
momenta of the nucleons and $s^{(\prime)}$ the polarizations. To obtain the
EDM, one  extrapolates the results for finite momentum transfer to
the point with $q^2=0$,
\begin{equation}
   d_N = \frac{F_{3,N}(0)}{2m_N}~.
\end{equation}
In Eq.~\eqref{eq:formfactor}, the CP-conserving parts are not shown, and  the
axial Gordon identity was used in the second step.
The latest results using the form factor method were reported in
Ref.~\cite{Shintani:2014talk}, where the calculation was performed on a lattice
with a volume of $(2.7~\text{fm})^3$, a lattice spacing of $a=0.11$~fm, and
pion masses of 330~MeV and 420~MeV. The smallest momentum transfer is about
$-q^2=0.2$~GeV$^2$. Various extrapolations or corrections are necessary in
order to obtain the result in the physical world: chiral extrapolation to the
physical pion mass, finite volume corrections, corrections due to the finite
volume spacing, and extrapolation from finite to zero momentum transfer. Chiral
perturbation theory (CHPT) is the proper theoretical framework to calculate
these corrections. A lot of work on the nucleon EDM in the framework of
CHPT has been done, see, e.g.,
Refs.~\cite{Crewther:1979pi,Pich:1991fq,Borasoy:2000pq,
O'Connell:2005un,Hockings:2005cn,Chen:2007ug, Narison:2008jp,
Ottnad:2009jw,Ottnadthesis,Mereghetti:2010tp,Mereghetti:2010kp,Bsaisou:2012rg,
Guo:2012vf, deVries:2012ab,Seng:2014pba,Dekens:2014jka}.

Finite volume corrections were considered before in
Refs.~\cite{O'Connell:2005un,Guo:2012vf}. However, in both works Lorentz
invariance was assumed to perform the tensor reduction of loop integrals. This
is not legitimate since the Lorentz symmetry is broken to the cubic symmetry on
a lattice with periodic boundary conditions, which is a torus. In this paper,
we will revisit this issue taking into account the breaking of Lorentz
symmetry. In fact, in this case, one cannot define the EDM as in the infinite
volume. Instead, we will calculate the finite volume corrections to the
CP-violating nucleon matrix elements of the electromagnetic current. The
calculations will be presented in Sec.~\ref{sec:fv}. The chiral extrapolation
of the neutron and proton EDMs will be discussed in Sec.~\ref{sec:ch}, and
Sec.~\ref{sec:sum} contains a brief summary.

\section{Finite volume corrections on a torus}
\label{sec:fv}

The decomposition of the nucleon matrix element of the electromagnetic current
in terms of form factors given in Eq.~\eqref{eq:formfactor} is based on the
Lorentz invariance and gauge symmetry. However, for a torus, the Lorentz
invariance is reduced to the cubic symmetry, and the decomposition is not valid
any more. In this section, we will evaluate the finite volume corrections to
the CP-violating part of the nucleon matrix
elements induced by the QCD $\theta$-term. For discussions on the finite volume
corrections
to the CP-conserving nucleon matrix elements, see, e.g.
Refs.~\cite{Tiburzi:2007ep,Jiang:2008ja,Greil:2011aa}.

\begin{figure}
    \centering
    \includegraphics[width=0.9\textwidth]{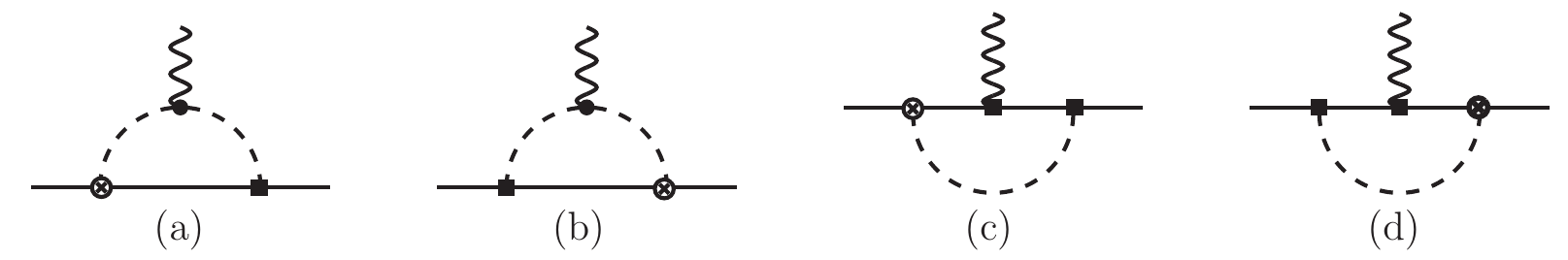}
    \caption{One-loop contributions to the nucleon EDFFs at NLO. Nucleons and
pions are represented by solid and dashed lines, respectively. $\otimes$, black
dots and filled squares denote CP-violating, second order mesonic and
first order meson-baryon vertices, respectively.}
    \label{fig:feyn}
\end{figure}

Finite volume corrections are a long-distance effect, and are dominated by the
degrees of freedom with the longest range. For our case, these corrections
are dominated by  pion-nucleon loops. The kaon-hyperon loops are suppressed relative to the
pion-nucleon loops by a factor of $e^{-(M_K-M_\pi)L}$, and thus will not be
considered here. Up to the next-to-leading order (NLO), the
one-loop diagrams contributing to the nucleon EDFFs
are shown in Fig.~\ref{fig:feyn}. Other one-loop diagrams contribute
from the  next-to-next-to-leading order in the chiral
expansion~\cite{Ottnad:2009jw,Ottnadthesis}. For the neutron, we need to
consider the $\{\pi^-,p\}$ loop, and for the proton, the loops of interest are
$\{ \pi^0,p\}$ and $\{\pi^+,n\}$.
The necessary Lagrangians for calculating these diagrams can be found
in Refs.~\cite{Borasoy:2000pq,Ottnad:2009jw,Ottnadthesis,Guo:2012vf}. A detailed
analysis at the one-loop level in the infinite volume can be found in these
references either, and we will focus on the finite volume corrections here.

We define the finite volume correction to a quantity $\mathcal{Q}$ as
\begin{equation}
    \delta_L[\mathcal{Q} ] = \mathcal{Q}(L) - \mathcal{Q}(\infty)~,
\end{equation}
where $\mathcal{Q}(L)$ and $\mathcal{Q}(\infty)$ denote the quantity in the
finite and infinite volumes, respectively. In the infinite volume, the contribution
from the pion loops to the CP-violating nucleon matrix element up to NLO is
given by
\begin{equation}
    \epsilon_\mu\left\langle N(p',s')| J^\mu | N(p,s) \right\rangle = i
\frac{8 e V_0^{(2)} \bar\theta_0}{F_\pi^4} \epsilon_\mu \bar u(p',s') \left[
C_\text{ab} \left(G_1^\mu(q)+G_2^\mu(q)\right) + C_\text{cd} G_3^\mu(q) \right]
\gamma_5 u(p,s)~,
\label{eq:me}
\end{equation}
where $e$ is the electric charge of the proton,
$V_0^{(2)}$ is a low-energy constant (LEC) of
U(3) chiral perturbation
theory~\cite{HerreraSiklody:1996pm}, $\epsilon_\mu$ is the
polarization vector of the photon, $\bar \theta_0$ is related to the measurable
vacuum angle $\theta_0$ via~\cite{Ottnad:2009jw}
\begin{equation}
 \bar\theta_0 = \left[ 1 + \frac{4 V_0^{(2)}}{F_\pi^2}
\frac{4 M_K^2 - M_\pi^2}{M_\pi^2\left(2 M_K^2 - M_\pi^2 \right)}\right]^{-1}
\,\theta_0 ~, 
\label{eq:theta0barMK}
\end{equation}
and the loop functions $G_1^\mu(q)$, $G_2^\mu(q)$ and $G_3^\mu(q)$ are given by
\begin{eqnarray}
    G_1^\mu(q) \al=\al i\! \int\! \frac{d^4k}{(2\pi)^4}
\frac{2k^\mu+q^\mu}{(k^2 - M_\pi^2) [(k+q )^2-M_\pi^2]}~, \nonumber\\
   G_2^\mu(q) \al=\al 2 m_N^2\, i\! \int\! \frac{d^4k}{(2\pi)^4}
\frac{2k^\mu+q^\mu}{(k^2 - M_\pi^2) [(k+q )^2-M_\pi^2] [(p-k)^2 - m_N^2
]}~,\nonumber\\
  G_3^\mu(q) \al =\al -4 m_N^2\, i\! \int\! \frac{d^4k}{(2\pi)^4}
\frac{k^\mu}{(k^2 - M_\pi^2) [(p-k )^2-m_N^2] [(p'-k)^2 - m_N^2
]}~.
\label{eq:Gi}
\end{eqnarray}
For the neutron matrix element, the coefficients $C_\text{ab}$ and
$C_\text{cd}$ are $2(D+F)(b_D+b_F)$ and $-2(D+F)(b_D+b_F)$, respectively, where
$D$ and $F$ are LECs in the leading order Lagrangian of baryon CHPT (the axial
coupling constants), and $b_D$
and $b_F$ are LECs in the NLO Lagrangian related to the baryon mass splittings.
For the proton,
$C_\text{ab}=-2(D+F)(b_D+b_F)$ originates from the $\{\pi^+,n\}$ loop and
$C_\text{cd}=-(D+F)(b_D+b_F)$ is generated from the $\{\pi^0,p\}$ loop. In the
following, we will work in the Lorentz gauge with $\epsilon_\mu q^\mu=0$.
Expressing Eq.~\eqref{eq:me} in terms of two-point and three-point scalar loop
functions in the infinite volume, we get~\cite{Guo:2012vf}
\begin{eqnarray}
  \al\al  C_\text{ab} \left(G_1^\mu(q)+G_2^\mu(q)\right) + C_\text{cd}
G_3^\mu(q) \nonumber\\
\al=\al (p+p')^\mu \!\left\{  C_\text{ab} \!\left[ - J_{MM}(q^2) +
\left(\! M_\pi^2 -\frac{q^2}{2} \right)\! J_{MMm}(q^2,m^2_N) \right]\! + \left(
C_\text{ab} + C_\text{cd} \right) J_{Mm}(m^2_N) \right\}~.\nonumber\\
\end{eqnarray}
The definitions and the analytic expressions using infrared regularization of
the loop functions $J_{MM}(q^2)$, $J_{Mm}(m^2_N)$ and $J_{MMm}(q^2,m^2_N)$ can be
found in App.~B of Ref.~\cite{Guo:2012vf}.

Next let us consider the finite volume corrections, which are due to the
quantization of the momentum on a torus. Assuming the temporal direction to be
infinite, we can integrate out the temporal component of the loop momentum by means
of contour integration. The integration over three-momentum will become a
sum, and the finite volume correction to a loop integral is
\begin{equation}
    \deltaL{G_i(q)} =  \left[ \frac1{L^3} \sum_{\vec{k}} -
\int\!\frac{d^3\vec{k} }{(2\pi)^3 } \right] I_i(k,q)~,
\end{equation}
where $I_i(k,q)$ denotes the integrand of the loop function $G_i(q)$. The finite
volume
corrections to the loops in Eq.~\eqref{eq:Gi} can be worked out as
\begin{eqnarray}
    \deltaL{G_1^\mu(q)} \al=\al - \frac{1}{2} \int_0^1\!dx 
\frac{\partial}{\partial q_{1i}} \sci_{1/2}( \Delta_1, \vec{q}_1 )~, \nonumber\\
     \deltaL{G_2^\mu(q)} \al=\al \frac32m_N^2 \int_0^1\!dx\int_0^1\!dy \,y
\bigg[ \frac13 \frac{\partial}{\partial q_{2i}}
\sci_{3/2}(\Delta_2, \vec{q}_2 ) + \bar y\,
p^\mu \,\sci_{5/2} (\Delta_2, \vec{q}_2) \bigg]~, \nonumber\\
    \deltaL{G_3^\mu(q)} \al=\al \frac32 m_N^2 \int_0^1\!dx\int_0^1\!dy \,y
\bigg[ \frac13 \frac{\partial}{\partial q_{3i}}
\sci_{3/2}(\Delta_3, \vec{q}_3 ) +  y\,
p^{\prime\,\mu}\, \sci_{5/2} (\Delta_3, \vec{q}_3) \bigg]~,
\label{eq:deltaGi}
\end{eqnarray}
where
\begin{align}
    \vec q_1 & = x \vec q, & \Delta_1 &= M_\pi^2 - x\bar x q^2~, &
\nonumber\\
    \vec q_2 & =  xy\vec q - \bar y \vec p, & \Delta_2 &= {\bar y}^2
m_N^2 + y M_\pi^2 - x\bar x y^2 q^2~, & \nonumber\\
    \vec q_3 & = xy \vec q - y\vec p\,', & \Delta_3 &= y^2 m_N^2 + \bar y
M_\pi^2 - x\bar x y^2 q^2~,&
\end{align}
with $\bar x=1-x$ and $\bar y=1-y$, see App.~\ref{app:fv} for definitions
and details. Therefore, the finite volume correction to
the CP-violating matrix element of the electromagnetic current for the neutron
is given by
\begin{equation}
    i \frac{16 e V_0^{(2)} \bar\theta_0}{F_\pi^4} C
    \big( \deltaL{G_1^\mu(q)} + \deltaL{G_2^\mu(q)} + \deltaL{G_3^\mu(q)} \big)
    \bar u(p',s') \gamma_5 u(p,s) ~,
    \label{eq:fvn}
\end{equation}
with $C=(D+F)(b_D+b_F)$. The correction to the proton matrix element is
\begin{equation}
    i \frac{16 e V_0^{(2)} \bar\theta_0}{F_\pi^4} C
    \left( -\deltaL{G_1^\mu(q)} - \deltaL{G_2^\mu(q)} +
\frac12\deltaL{G_3^\mu(q)} \right)
    \bar u(p',s') \gamma_5 u(p,s) ~.
    \label{eq:fvp}
\end{equation}
From Eqs.~\eqref{eq:fvn} and \eqref{eq:fvp}, it is clear that the matrix
elements on a torus cannot be written in the form of Eq.~\eqref{eq:formfactor}
as a consequence of the lack of Lorentz symmetry.

In order to investigate how large the finite volume correction is, we
calculate the ratio of the correction over the infinite volume result of
the matrix element which can be defined as, see Eq.~\eqref{eq:formfactor},
\begin{equation}
    \tilde F_3^\mu(p,q) = \frac{F_3(q^2)}{2 m_N} (p+p')^\mu.
\end{equation}
The infinite volume expressions for both the neutron and proton EDFFs were
calculated in Refs.~\cite{Ottnad:2009jw,Ottnadthesis,Guo:2012vf}, and they are
given in App.~\ref{app:infinite} for completeness.
The ratios for the neutron and proton are given by
\begin{equation}
    R_n^\mu = \frac{16 e V_0^{(2)}\bar\theta_0}{ F_\pi^4 \tilde F_{3,n}^\mu }
\big( \deltaL{G_1^\mu(q)} + \deltaL{G_2^\mu(q)} + \deltaL{G_3^\mu(q)} \big)~,
\end{equation}
and
\begin{equation}
    R_p^\mu = \frac{16 e V_0^{(2)}\bar\theta_0}{ F_\pi^4 \tilde F_{3,p}^\mu }
\big( 2\, \deltaL{G_1^\mu(q)} + 2\,\deltaL{G_2^\mu(q)}  -
\deltaL{G_3^\mu(q)} \big)~,
\end{equation}
respectively.

\begin{figure}[t]
    \centering
    \includegraphics[width=0.48\textwidth]{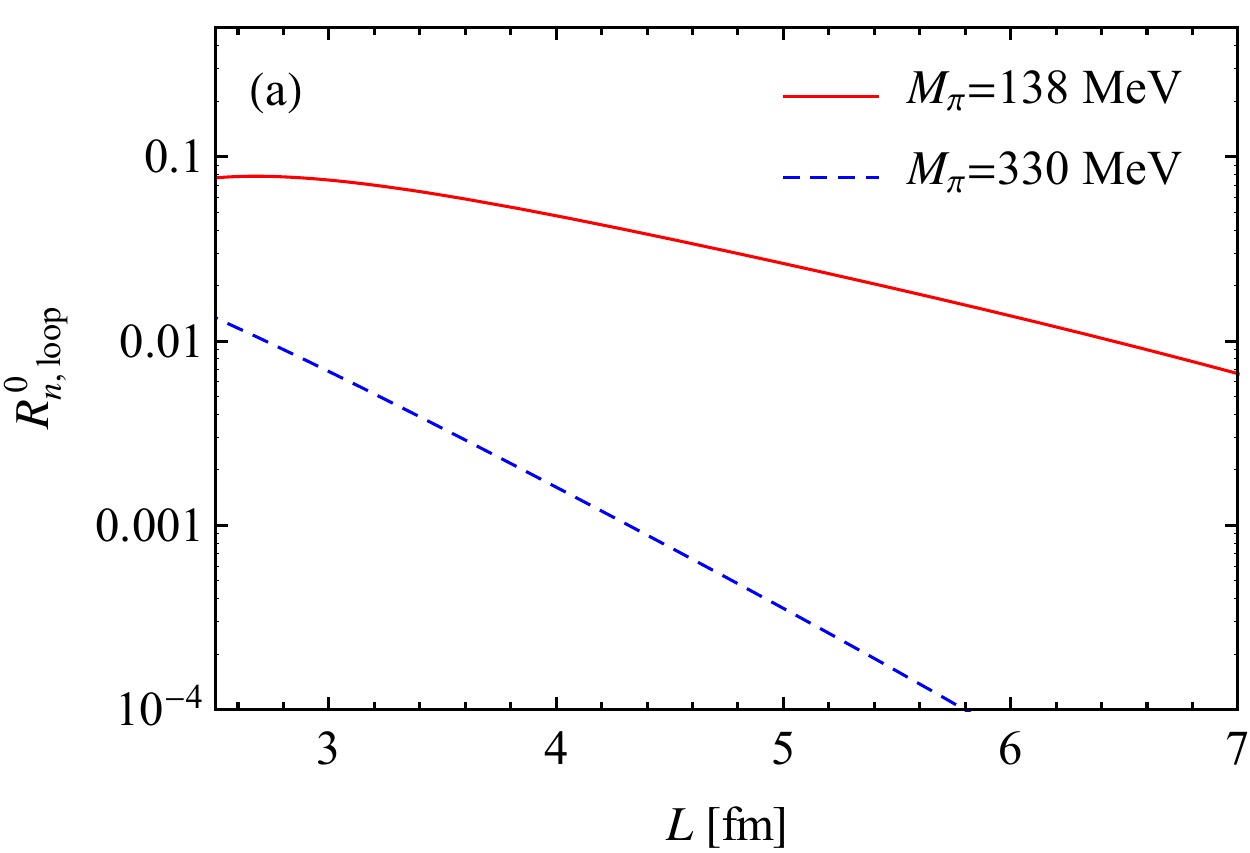}\hfill
    \includegraphics[width=0.48\textwidth]{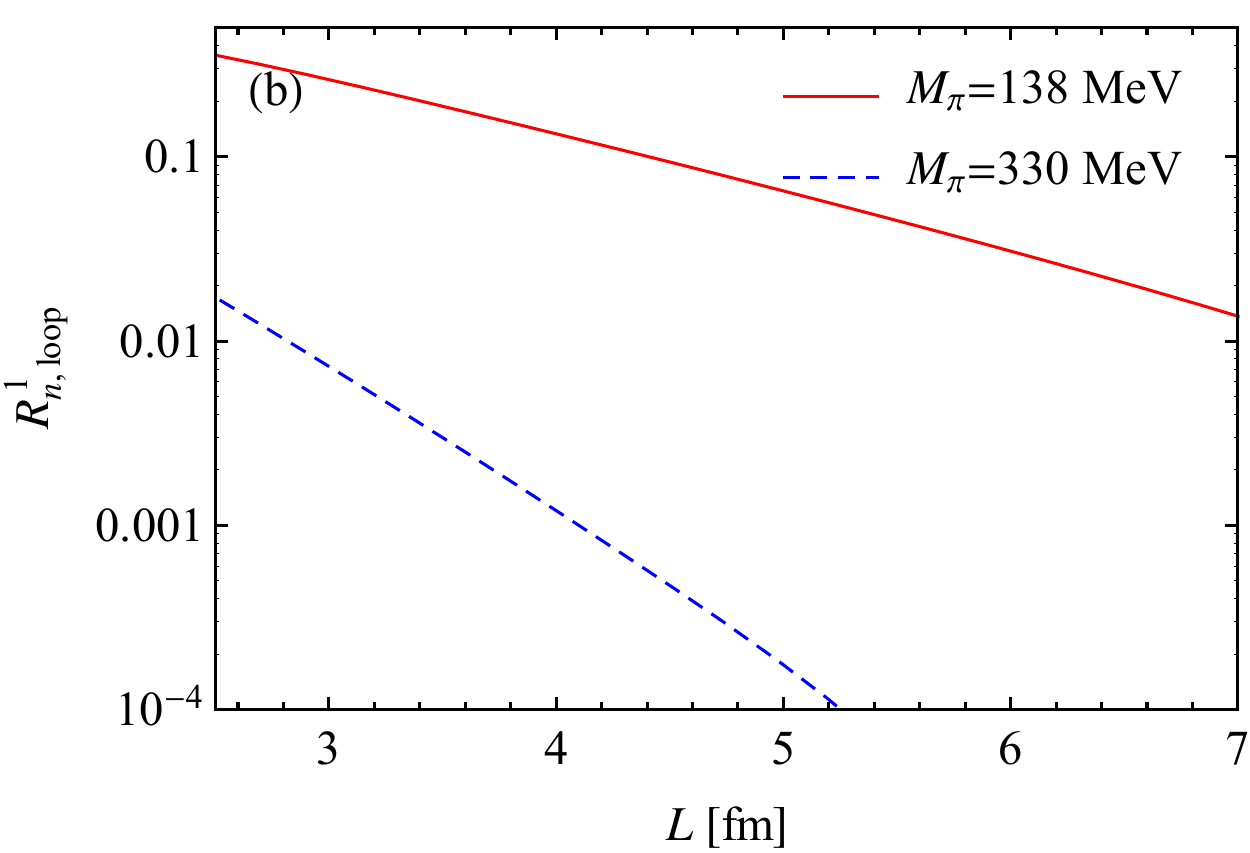}\\[3mm]
    \includegraphics[width=0.48\textwidth]{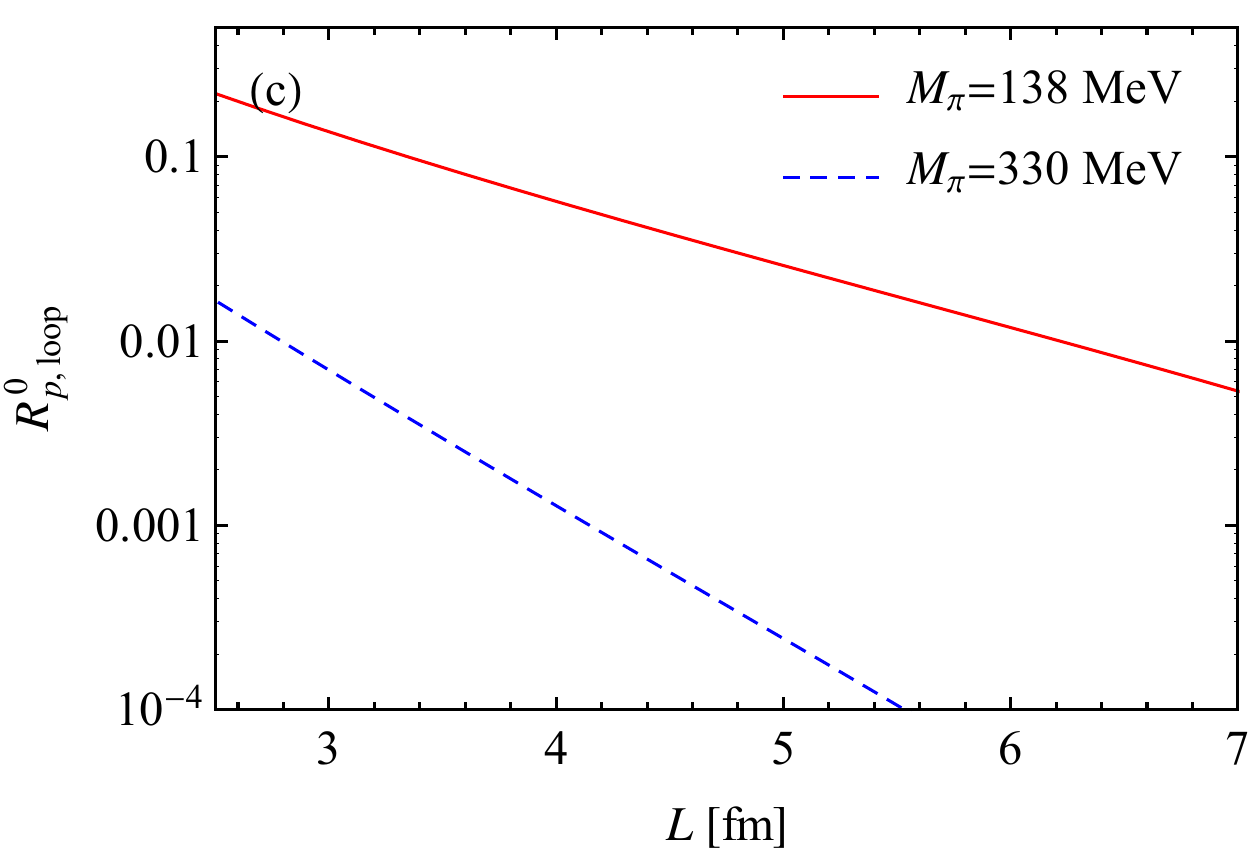}\hfill
    \includegraphics[width=0.48\textwidth]{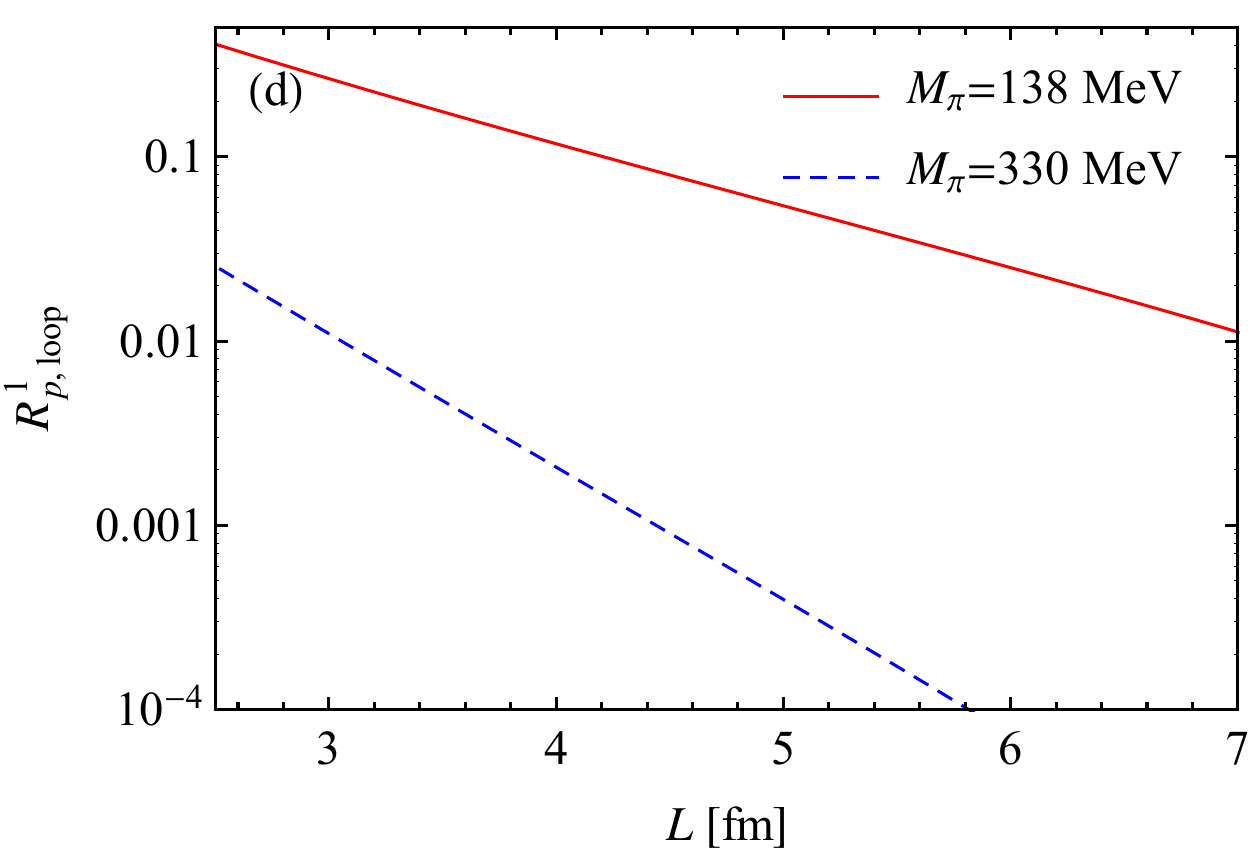}
    \caption{Ratios of the finite volume corrections to the loop  contributions
to the CP-violating nucleon matrix elements of the electromagnetic current. The
three-momenta are $\vec p=\{-2\pi/L,0,0\}$ and $\vec q=\{2\pi/L,0,0 \}$. (a)
and (b) are for the temporal and first spatial components for the neutron,
respectively, while (c) and (d) are for the proton. }
    \label{fig:ratios}
\end{figure}

For numerical values, we take $F_\pi=92.2$~MeV~\cite{Beringer:1900zz},
$V_0^{(2)}\simeq-5\times10^{-4}$~GeV$^4$~\cite{HerreraSiklody:1997kd},
$D=0.804$ and $F=0.463$~\cite{Goto:1999by}. From a leading order fitting to the
octet baryon mass differences, we get $b_D=0.068$~GeV$^{-1}$ and
$b_F=-0.209$~GeV$^{-1}$.
The  pion mass dependence of the pion decay constant, the nucleon mass and the mass
of the eta meson will be neglected since they contribute from the
next-to-next-to-leading order. As mentioned before, because the Lorentz
symmetry is broken, the finite volume corrections
do not only depend on $q^2$. Different $q^\mu$ with the same $q^2$ can result
in different corrections, as pointed out in, e.g. Ref.~\cite{Tiburzi:2007ep}.
In lattice calculations of form factors, in order to reduce the statistical
noise, often  the momentum of the sink is set to zero and the
momenta of the source and the current take small values. The cubic symmetry of
a torus ensures the equivalence of the three spatial directions. Thus, we
take $\vec p = \{ -2\pi/L,0,0 \}$ and $\vec q = \{ 2\pi/L, 0, 0\}$ to show the
numerical results of the ratios defined above. In this case, as can be seen from
the expressions, only the temporal and the first spatial components of the
matrix elements, both in infinite and finite volumes, are nonvanishing.
The ratios for both the neutron and proton are shown in
Fig.~\ref{fig:ratios}. The solid curves are the results with a physical pion
mass, and the dashed ones are for $M_\pi=330$~MeV, which is the smallest pion
mass used in the recent lattice calculation~\cite{Shintani:2014talk}. Notice
that in the plots, we have neglected the contribution from the counterterms
$w_a$ and $w_b$  and the regularization scale in the infinite volume matrix
elements is taken as $\mu = 1$~GeV. That means that we compare the loop contribution
in the finite to the one in the infinite volume at a given natural scale (as
indicated by the subscript 'loop' in the figure). This is done for better
displaying of the corrections as the tree contributions to the nucleon EDM
contain some sizeable uncertainties as discussed in the next section.
One sees that the correction to the spatial
matrix element is larger than that to the temporal one.  In the case of a 330~MeV
pion mass, the finite volume corrections  for $L\geq2.5$~fm are always smaller
than $3\%$ of the loop contributions for both the neutron and the proton. In the
case of the physical mass, in order to have a correction less than $10\%\,
(5\%)$ for the neutron, $L$ needs to be larger than $4.4\,(5.4)$~fm. The values
for the proton are similar.

\section{Chiral extrapolation of the neutron and proton EDMs}
\label{sec:ch}

\begin{figure}[t]
    \centering
\includegraphics[width=0.48\textwidth]{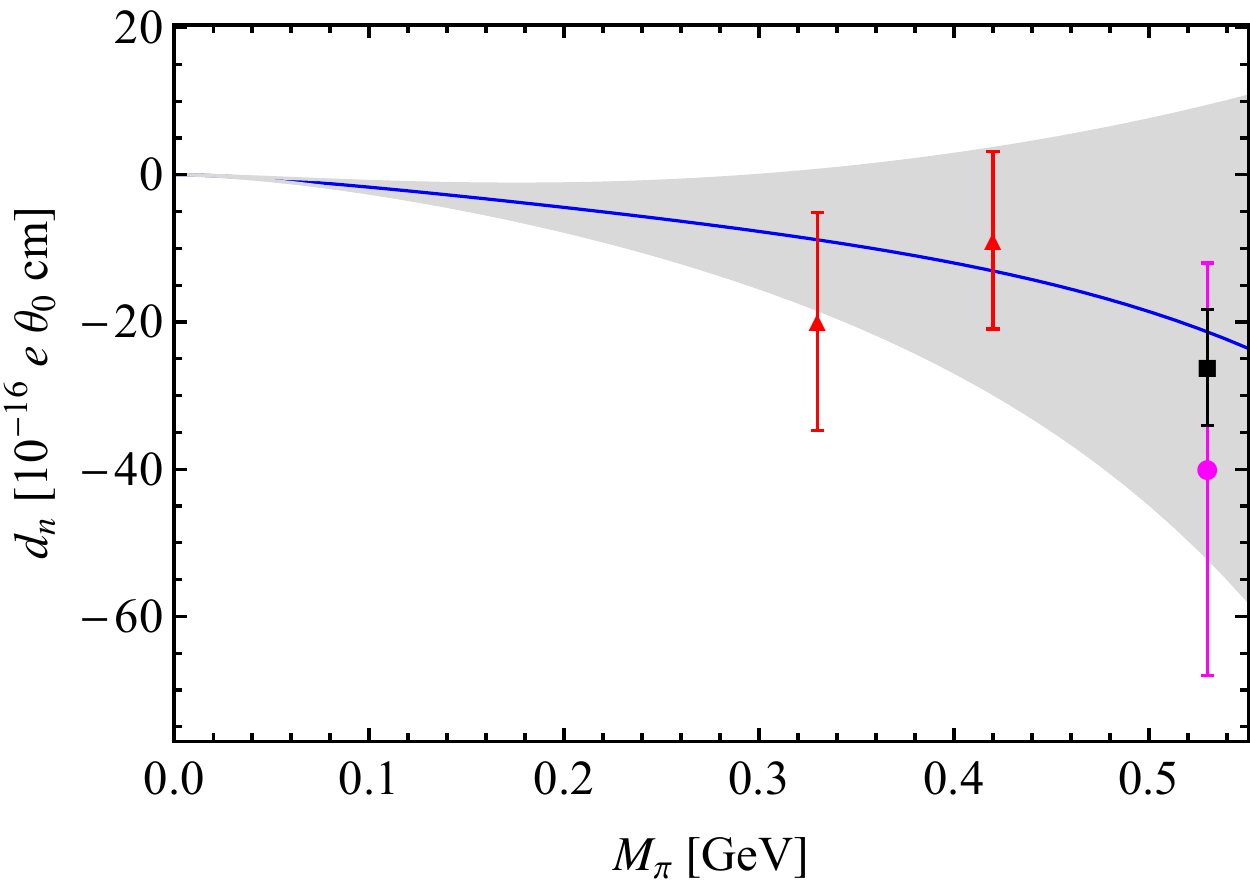}\hfill
    \includegraphics[width=0.48\textwidth]{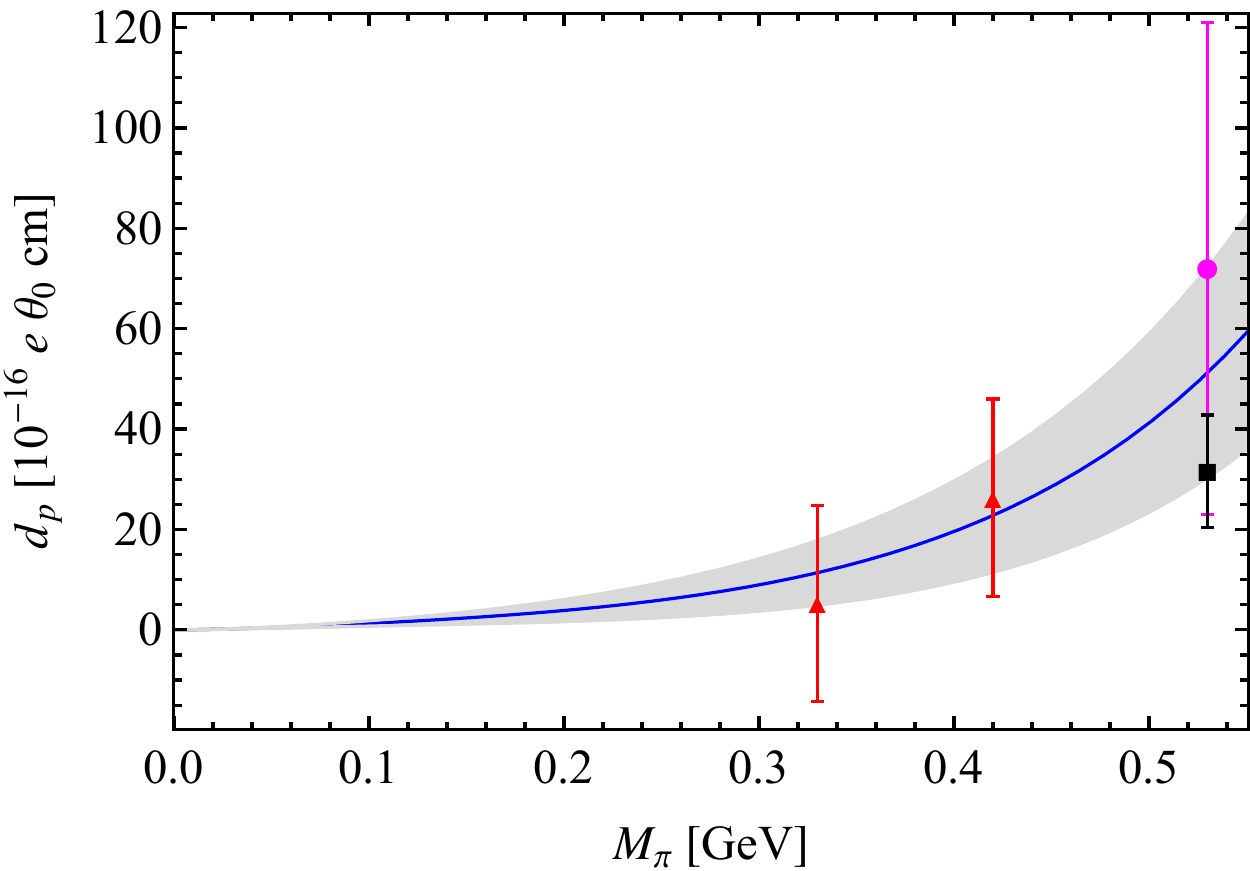}
    \caption{Fit to the lattice results of the neutron and proton EDMs at
$M_\pi=330$~MeV and $420$~MeV~\cite{Shintani:2014talk} (filled triangles with
error bars) with the two counterterms $w_a(\mu)$ and $w_b(\mu)$.
The filled circle and square with uncertainties at $M_\pi=530$~MeV are the
lattice results reported in Refs.~\cite{Shintani:2008nt}
and~\cite{Shintani:2012talk}, respectively.}
    \label{fig:fit}
\end{figure}

As stressed in Ref.~\cite{Guo:2012vf}, the baryon octet EDMs depend on two
LECs at NLO, which are called $w_a$ and $w_b$.
So far the only knowledge of the values of the LECs $w_a(\mu)$ and
$w_b(\mu)$ is from a determination~\cite{Guo:2012vf} using the lattice data for
the neutron and proton EDMs at a large pion mass of
530~MeV~\cite{Shintani:2012talk}. Here we will use the updated lattice results
at smaller pion masses of 330 and 420~MeV~\cite{Shintani:2014talk} for  a
new determination. As shown in the last section, the finite volume corrections
to the relevant nucleon matrix elements at $M_\pi=330$~MeV are smaller than
$3\%$, which is much smaller than the error bars of the lattice data. Those at
the larger pion mass of 420~MeV are even smaller. Therefore, we can neglect the
finite volume effect, and fit to the lattice data at both pion masses using the
expressions in Eqs.~\eqref{eq:F3n} and~\eqref{eq:EDFFp} with $q^2=0$ given in
App.~\ref{app:infinite}.
Fig.~\ref{fig:fit} presents the best fit with the bands reflecting the
uncertainties propagated from those of the lattice data. The scale is taken to
be $\mu=1$~GeV. We obtain
\begin{equation}
    w_a(1~\text{GeV}) = (-0.02\pm0.04)~\text{GeV}^{-1},\qquad w_b(1~\text{GeV})
= (-0.32\pm0.05)~\text{GeV}^{-1}~.
\label{eq:wawb}
\end{equation}
Using these values, the neutron and proton EDMs at the physical pion mass are
predicted to be  (in units of $10^{-16} e\, \theta_0\,$cm)
\begin{equation}
    d_n = -2.7\pm0.8\pm0.8, \qquad d_p = 2.1\pm0.6\pm1.0~,
    \label{eq:dndp}
\end{equation}
where the first uncertainties are from the counterterms $w_a$ and $w_b$, and
the second are from varying $\mu$ between the mass of the rho meson and the
mass of the $\Xi$. Physical observables do not depend on the choice of the
regularization scale, and the scale dependence of the loops is cancelled by
that of the counterterms. However, when we choose to use the values of the
LECs at a given scale, we may vary the scale in the loops within a
certain range. Such a variation is a higher order effect, and we thus use it to
estimate the uncertainties due to neglecting higher order contributions.
If we only consider the loops, then we obtain~\cite{Guo:2012vf}
\begin{equation}
    d_n^\text{loop} = -3.1\pm0.8, \qquad d_p^\text{loop}=5.6\pm1.0
\end{equation}
in units of $10^{-16} e\, \theta_0\,$cm. Comparing with the values in
Eq.~\eqref{eq:dndp}, one sees that the
neutron EDM is dominated by the loops while the counterterms and the loop
contribution to the proton EDM are of similar size at  $\mu=1$~GeV.

\section{Summary}
\label{sec:sum}

In this paper, we have calculated the finite volume corrections to the
CP-violating nucleon matrix elements of the electromagnetic current on a
torus taking into account the breaking of the Lorentz symmetry. The matrix
elements do not depend only on $q^2$, which is the case in the continuum, and
the corrections for different space-time directions
differ from each another. We used two pion
masses to investigate the size of the corrections explicitly. For the pion mass
of 330~MeV, which is the lowest pion mass employed in the lattice
calculation reported in~\cite{Shintani:2014talk}, we found that the finite
volume corrections are negligible for $L\geq2.5$~fm in comparison with the
other uncertainties of the lattice results. When the pion takes its physical
mass, the corrections can be significant. In order to have a correction smaller
than $10\%$, one needs a lattice size $\gtrsim4$~fm for the neutron and
$\gtrsim6$~fm for the proton (this estimate is based on infinite volume results
including both loops and counterterms, and the central values of the
counterterms $w_a$ and $w_b$ given in Eq.~\eqref{eq:wawb} are used here).
We also performed a chiral extrapolation of the neutron and proton EDMs of the
lattice results~\cite{Shintani:2014talk}, and obtain
$d_n=(-2.7\pm1.2)\times10^{-16}e\,\theta_0\,$cm and
$d_p=(2.1\pm1.2)\times10^{-16}e\,\theta_0\,$cm. In the future, it might
also be interesting to consider finite volume corrections with
twisted boundary conditions, which would be helpful to reduce the systematic
uncertainties of lattice calculations due to the momentum extrapolation.

\medskip

\section*{Acknowledgments}
We are grateful to E.~Shintani for providing us their lattice data
and for valuable discussion. We would like to thank
A.~J\"{u}ttner for helpful discussion and T.~Greil for some useful
communications.
This work is supported in part by the DFG  and the NSFC
through funds provided to the Sino-German CRC 110 ``Symmetries and the Emergence
of Structure in QCD'', by the EU I3HP ``Study of Strongly Interacting Matter''
under the Seventh Framework Program of the EU, and by the NSFC (Grant No.
11165005).

\medskip

\begin{appendix}

\section{Formulae for finite volume corrections}
\label{app:fv}

All the finite volume corrections can be written in terms of corrections to
loop integrals. The basic function is given
by~\cite{Sachrajda:2004mi,Jiang:2008ja} (here, $m$ denotes a quantity with
dimension mass)
\begin{eqnarray}
    \sci_\beta(m^2,\vec A\,) \al\equiv\al
\left( \frac1{L^3}\sum_{\vec k} - \int\!\frac{d^3\vec k}{(2\pi)^3 } \right)
\frac1{ \left[(\vec k +\vec A\,)^2 + m^2 \right]^{\beta } }
\nonumber\\
 \al=\al \frac1{(4\pi)^{3/2} \Gamma(\beta)} \int_0^\infty \!d\tau \,
\tau^{\beta-\frac52 } e^{-\tau m^2 } \left[ \prod_{i=1}^3 \vartheta_3 \left(
\frac{A_iL }{2}, e^{ - \frac{L^2}{4\tau} } \right) - 1 \right]~,
\end{eqnarray}
where
\begin{equation}
    \vartheta_3(z,q) \equiv \sum_{n=-\infty}^\infty q^{(n^2)} e^{i 2 nz } = 1 +
\sum_{n=1}^\infty 2\, q^{(n^2)} \cos(2nz)
\end{equation}
is the Jacobi elliptic theta function. This function is even in $\vec A$, i.e.
$\sci_\beta(m,\vec A\,)=\sci_\beta(m,-\vec A\,)$. For a large value of $L$, one
has the asymptotic form
\begin{equation}
\sci_\beta (m^2,\vec A\,) \,{\rightarrow}\, \frac{e^{ - mL}} {4\pi
m\Gamma(\beta)} \left( \frac{2m}{L} \right)^{2-\beta} \left[
\sum_{i=1}^3 \cos(A_i L) \right]~.
\end{equation}
Thus, we see that the finite volume correction is exponentionally suppressed.

When there is a factor of momentum in the numerator of the loop integrand,
using
\begin{equation}
    \frac{\partial}{\partial A^i } \left[ (\vec k +\vec A\,)^2 + m^2
\right]^{-\beta } = -\frac{2\beta(k^i+A^i) }{ \left[(\vec k +\vec A\,)^2+m^2
\right]^{\beta+1} }~,
\end{equation}
one can get the following relations~\cite{Jiang:2008ja,Greil:2011aa}
\begin{eqnarray}
    \sci_\beta^i(m^2,\vec A\,) \al\equiv\al
\left( \frac1{L^3}\sum_{\vec k} - \int\!\frac{d^3\vec k}{(2\pi)^3 } \right)
\frac{k^i}{ \left[(\vec k +\vec A\,)^2 + m^2 \right]^{\beta } }
\nonumber\\
 \al=\al - A^i \sci_{\beta}(m^2,\vec A\,) -
\frac1{2(\beta-1) } \frac{\partial}{\partial A_i} \sci_{\beta-1}(m^2,\vec A\,)~,
\nonumber\\
    \sci_\beta^{ij}(m^2,\vec A\,) \al\equiv\al
 \left( \frac1{L^3}\sum_{\vec k} - \int\!\frac{d^3\vec k}{(2\pi)^3 } \right)
\frac{k^i k^j}{ \left[(\vec k +\vec A\,)^2 + m^2 \right]^{\beta } }
\nonumber\\
 \al = \al  A^i A^j \sci_{\beta}(m^2,\vec A\,) +
\frac1{2(\beta-1)} \left( A^i \frac{\partial}{\partial A_j} + A^j
\frac{\partial}{\partial A_i} + \delta^{ij} \right)  \sci_{\beta-1}(m^2,\vec
A\,)  \nonumber\\
\al\al + \frac1{4(\beta-1)(\beta-2)} \frac{\partial^2}{\partial
A_i \partial A_j} \sci_{\beta-2}(m^2,\vec A\,) ~.
\end{eqnarray}

\section{Nucleon electric dipole form factors in infinite volume}
\label{app:infinite}

The NLO expressions for the EDFFs of the nucleons in U(3) CHPT has been worked
out in Refs.~\cite{Ottnad:2009jw,Ottnadthesis,Guo:2012vf}. They depend on two
counterterms $w_a(\mu)$ and $w_b(\mu)$ which are combinations of several LECs in
the meson and baryon chiral Lagrangians~\cite{Guo:2012vf}. The neutron EDFF
reads
\begin{eqnarray}
   \frac{F_{3,n}(q^2)}{2 m_N} \al=\al \frac83  w_a(\mu)e \bar{\theta }_0
   + \frac{V_0^{(2)} e \bar\theta_0}{\pi^2 F_\pi^4} \Bigg\{
   (D+F)\left(b_D+b_F\right) I_\pi - (D-F)\left(b_D-b_F\right) I_K
    \nonumber\\
    \al\al +  8 (D-F) \left(b_D-b_F\right)^2 \left(M_K^2-M_\pi^2\right)
\frac{\pi}{\sqrt{-q^2}}   \arctan \frac{\sqrt{-q^2}}{2 M_K} \Bigg]
\Bigg\}~,
   \label{eq:F3n}
\end{eqnarray}
where
\[
I_{\pi(K)} = 1 - \ln \frac{M_{\pi(K)}^2}{\mu^2}
   + \sigma_{\pi(K)} \ln\frac{\sigma_{\pi(K)}-1}{\sigma_{\pi(K)}+1}  +
   \frac{\pi \left(2M_{\pi(K)}^2-q^2\right)} {2m_N\sqrt{-q^2}} \arctan
   \frac{\sqrt{-q^2}}{2M_{\pi(K)}}
\]
with $\sigma_{\pi(K)}=\sqrt{1-4M_{\pi(K)}^2/q^2}$, and the proton EDFF is given
by
\begin{eqnarray}
   \frac{F_{3,p}(q^2)}{2 m_N} \al=\al -\frac43 e \bar{\theta }_0
\left[w_a(\mu) + w_b(\mu) \right] - \frac{V_0^{(2)} e \bar\theta_0}{6\pi^2
F_\pi^4} \Bigg\{ 6
   (D+F) \left(b_D+b_F\right)
    \Bigg( I_\pi + \frac{3\pi M_\pi}{2 m_N} \Bigg) \nonumber\\
    \al\al + 4 \left(Db_D + 3Fb_F\right) \left( I_K + \frac{\pi M_K}{m_N}
\right) \nonumber\\
   \al\al  + 32 \left(M_K^2-M_\pi^2\right) \left[ F \left(b_D^2+3
b_F^2\right)-\frac23 D
   b_D \left(b_D-3 b_F\right) \right] \frac{\pi}{\sqrt{-q^2}}
   \arctan\frac{\sqrt{-q^2}}{2 M_K} \nonumber\\
   \al\al + \frac{\pi}{m_N} \bigg[ 6 (D-F) \left(b_D-b_F\right) M_K + (D-3F)
   \left(b_D-3b_F\right) M_{\eta_8} \bigg] \Bigg\}~.
    \label{eq:EDFFp}
\end{eqnarray}

\end{appendix}

\end{document}